# Cooperative Parallel Particle Filters for online model selection and applications to Urban Mobility

L. Martino⋆, J. Read⊤, V. Elvira⋄, F. Louzada⋆

⋆ Universidade de São Paulo, São Carlos (Brazil).
⊤ Télécom ParisTech, Université Paris-Saclay. (France),
⋄ Universidad Carlos III de Madrid, Leganés (Spain).

**Abstract**

We design a sequential Monte Carlo scheme for the dual purpose of Bayesian inference and model selection. We consider the application context of urban mobility, where several modalities of transport and different measurement devices can be employed. Therefore, we address the joint problem of online tracking and detection of the current modality. For this purpose, we use interacting parallel particle filters, each one addressing a different model. They cooperate for providing a global estimator of the variable of interest and, at the same time, an approximation of the posterior density of each model given the data. The interaction occurs by a parsimonious distribution of the computational effort, with online adaptation for the number of particles of each filter according to the posterior probability of the corresponding model. The resulting scheme is simple and flexible. We have tested the novel technique in different numerical experiments with artificial and real data, which confirm the robustness of the proposed scheme.

## 1 Introduction

Monte Carlo (MC) algorithms are very popular numerical techniques for the approximation of optimal a posteriori estimators [9, 18, 13, 30]. Particle filters (PFs) are well-known MC methods that have been extensively applied in different fields, in order to handle analytically intractable posterior probability density functions (pdfs) [4, 10, 15, 20]. In this work, we consider the problem of tracking a variable of interest within a state-space model, where the dynamic and observation equations are unknown. A finite set of candidate models is considered, and the true model may be included (or not) within this set. Thus, the goal is both sequential tracking and online model selection [14, 24]. This problem is particularly of interest in an urban mobility context where different modalities of transport can be employed (e.g., bus, train, cycling). In the following, we discuss the main contributions of this work, the range of their applicability, and also related work.

**Contributions and organization of the paper.** In this work, we present a simple approach involving parallel particle filters (PFs), called *model averaging parallel particle filters* (MAPF), based on the Bayesian model averaging (BMA) principle [21]. The proposed PF solution performs the inference by running as many filters as candidate models, i.e., each PF is tailored to a different states-space model (Section 4). The parallel PFs in MAPF cooperate to provide a global approximation of the posterior distribution of the variable of interest given the data and, at the same time, also provide particle approximations of the posterior distributions of the models given the data. The interaction among the filters occurs by a dynamic allocation of the computational effort, i.e., distributing a portion of the total number of particles to each filter, proportionally to the posterior pdf of the corresponding model. Namely, the parallel filters in MAPF exchange information adapting online the number of particles employed in each filter. However, the total number of particles remains fixed, chosen in advance by the user. Hence, the novel scheme is able to distribute the computational effort online, rewarding the filters which are considering the most probable models given the observed data.

An exhaustive theoretical derivation of MAPF is provided in Sections 2-3. In Section 2, we introduce the general Bayesian formulation for tackling our problem. The posterior distribution which we study is *doubly* intractable, in the sense that it cannot be evaluated analytically and the computation of the related moments is analytically intractable. Then, the importance sampling (IS) technique for approximating the corresponding optimal Bayesian estimators is described in Section 3. The sequential IS approach is presented in Section 3.1. The use of parallel particle filters naturally appears following the theoretical derivation of MAPF. We have focused special attention in describing the sequential approximation of the marginal likelihood (a.k.a., Bayesian evidence), in Sections 3.1, 3.2 and further material is also given in different appendices. Different possible formulations are derived and discussed in detail, thus providing a concise review which could be of value to interested practitioners and researchers. Interesting special cases of MAPF are discussed in Section 4.2. For instance, when all the candidate models share the same likelihood function, MAPF can be seen as a unique PF with adaptive prior density. Moreover, when all the candidate models are equal and coincide with the true model, then MAPF can be interpreted as a distributed PF scheme where the filters compete in order to obtain more particles (according to the performance at each specific run). The application of MAPF in the case of time-varying models is described in Section 5.

We test MAPF in different experimental scenarios and apply the proposed scheme in a real data problem: an urban mobility context for detecting different modes of activity. For instance, one possible goal in urban mobility is to detect if a traveller is walking or riding a bus (or switching between these two modalities). The numerical results in a Section 6, with both synthetic and real data, show the efficiency and flexibility of the proposed algorithms.[1]

**Range of applicability.** The range of applicability of MAPF is wide, clearly not restricted only to the urban mobility case. Application of sequential model selection problem are very common in different fields where a stream of data is observed [11, 14, 23, 24]. For instance, in financial analysis where price fluctuations act differently under different regimes, as the ongoing instability in times of crisis [34]. They have also been used for fraud detection [22], explicitly modeling the switch to a regime of fraudulent activity from ordinary activity. In the medical domain, patients may need to be modeled for possible complications [37]. In marine tracking [38], vessels must be suitable monitored depending on different conditions of the environment. Another general application, where MAPF can be also applied, is the change detection and system identification problem [3]. Furthermore, MAPF can be used for optimizing the tuning of one or several parameters (problem specifically tackled in [2, 35]), selecting them within of a set of possible candidates (e.g., see Section 6.2).

---

[1] A preliminary MATLAB code of MAPF is provided at https://www.mathworks.com/matlabcentral/fileexchange/58597-model-averanging-particle-filter.

**Related works.** The problem addressed in this work is related to the *switching model* [11, 36, 31, 7] and the *multiple model* approaches [1, 5, 11]. The first main difference is that, in general, this kind of methods requires the definition and tuning of a transition probability matrix among the models (formed by, at least, $K^2$ parameters). MAPF is a simpler scheme since the transition matrix is not required so that the design effort (in terms of construction and tuning) can be focused completely on a proper choice of the dynamic and likelihood functions.[2] In a time-varying model scenario, MAPF only requires the tuning of an additional scalar parameter (denoted as $T_V$ in the rest of the work). We also remark that, in a standard tracking and model selection setup (when the true model is unknown and is not varying with the time), this additional scalar parameter is even not needed. Another main difference is that MAPF does not allow exchange of particles among the filters (with the exception of the refreshing steps; see Section 5), whereas in the switching model approach this exchange is usually allowed. This difference yields that MAPF can provide more accurate estimates (taking also advantage of the parallelization), since each filter employs its own particles without hybrid mixes which can jeopardize the tracking. Moreover, MAPF automatically adapts the number of particles of each PF keeping fixed the overall computational cost. PF schemes with adaptive number of particles has been proposed in literature [19, 17], but in the context of a unique filter. Moreover unlike in these works, in MAPF, the overall computational cost is kept constant, chosen in advance by the user. Other related and well-known approaches for the joint purpose of sequential tracking and parameter estimation are given in [2, 35]. In [14], the Kolmogorov-Smirnov test is applied for testing a dynamic equation in a state-space model. Finally, it is important to mention that specific particle filters addressing models with unknown statistics have been also designed [15, 32].

## 2 Bayesian inference and model selection in state-space models

Let us denote the unknown state $\mathbf{x}_t \in \mathcal{X}$ with $\mathcal{X} \in \mathbb{R}^{d_x}$ (continuous space) or $\mathcal{X} \in \mathbb{N}^{d_x}$ (discrete space), $t \in \mathbb{N}$, and the current observed data as $\mathbf{y}_t \in \mathcal{Y} \in \mathbb{R}^{d_y}$. We assume that the hidden sequence $\mathbf{x}_{1:T} = [\mathbf{x}_1, \ldots, \mathbf{x}_T]$ is generated with a transition pdf $g(\mathbf{x}_t|\mathbf{x}_{t-1})$. At the $t$-th iteration, we observe $\mathbf{y}_t$ with probability $f(\mathbf{y}_t|\mathbf{x}_t)$, so that after $T$ iterations we have $\mathbf{y}_{1:T} = [\mathbf{y}_1, \ldots, \mathbf{y}_T]$. The previous two pdfs, $g_t$ and $f_t$, jointly compose the true model indicated as $\mathcal{T}$, which we consider unknown. Namely, setting $g_1(\mathbf{x}_1|\mathbf{x}_0) = g_1(\mathbf{x}_1)$, we have

$$\mathcal{T}: \begin{cases} g_t(\mathbf{x}_t|\mathbf{x}_{t-1}) \\ f_t(\mathbf{y}_t|\mathbf{x}_t) \end{cases}, \qquad t = 1, \ldots, T. \tag{1}$$

---

[2]We remark that, in MAPF, every model can differ from the other ones for both dynamic and measurement equations.

For the sake of simplicity, we consider the model $\mathcal{T}$ fixed over the time $t$, however the case of time-varying model $\mathcal{T}_t$ is also tackled in Section 5.

We are interested in inferring the hidden states $\mathbf{x}_{1:T}$ given all the observed measurements $\mathbf{y}_{1:T}$. Both, $\mathbf{x}_{1:T}$ and $\mathbf{y}_{1:T}$, are generated according to the model $\mathcal{T}$ in Eq. (1). Since $\mathcal{T}$ is unknown, we consider $K$ different possible models, denoted as $\mathcal{M}_k$, with $k \in \{1, \ldots, K\}$, formed by a transition pdf and a likelihood function, i.e., setting $q_{k,1}(\mathbf{x}_1|\mathbf{x}_0) = q_{k,1}(\mathbf{x}_1)$,

$$\mathcal{M}_k : \begin{cases} q_{k,t}(\mathbf{x}_t|\mathbf{x}_{t-1}) \\ \ell_{k,t}(\mathbf{y}_t|\mathbf{x}_t) \end{cases}, \qquad t = 1, \ldots, T. \tag{2}$$

We denote the set of all considered models as $\mathcal{F} = \{\mathcal{M}_1, \ldots, \mathcal{M}_K\}$ The true model $\mathcal{T}$ could be contained in $\mathcal{F}$, i.e., $\mathcal{T} \in \mathcal{F}$, but in general we have $\mathcal{T} \notin \mathcal{F}$. However, even in the case $\mathcal{T} \notin \mathcal{F}$, we apply the Bayesian model averaging (BMA) approach [21] which provides a coherent mechanism for taking in account the model uncertainty, improving the overall filtering performance.

We assume a prior probability mass function (pmf), $p(\mathcal{M}_k)$, $k = 1, \ldots, K < \infty$ over the different models. Thus, the goal is to make inference about the sequence $\mathbf{x}_{1:T}$ and the $K$ different possible models, given the set of received measurements $\mathbf{y}_{1:T}$. Therefore, we study the following posterior density

$$p(\mathbf{x}_{1:T}|\mathbf{y}_{1:T}) = \sum_{k=1}^{K} p(\mathbf{x}_{1:T}, \mathcal{M}_k|\mathbf{y}_{1:T}), \tag{3}$$

$$= \sum_{k=1}^{K} p(\mathbf{x}_{1:T}|\mathbf{y}_{1:T}, \mathcal{M}_k) p(\mathcal{M}_k|\mathbf{y}_{1:T}), \tag{4}$$

$$= \frac{1}{p(\mathbf{y}_{1:T})} \sum_{k=1}^{K} p(\mathbf{x}_{1:T}|\mathbf{y}_{1:T}, \mathcal{M}_k) p(\mathbf{y}_{1:T}|\mathcal{M}_k) p(\mathcal{M}_k), \tag{5}$$

where $p(\mathbf{y}_{1:T}) = \sum_{j}^{K} p(\mathbf{y}_{1:T}|\mathcal{M}_j) p(\mathcal{M}_j)$. In general, the study of the posterior pdf above is (doubly) analytically intractable because of the following reasons. On the one hand, we cannot evaluate $p(\mathbf{x}_{1:T}|\mathbf{y}_{1:T})$ in Eq. (5) completely, since we cannot evaluate

$$p(\mathcal{M}_k|\mathbf{y}_{1:T}) = \frac{Z_{k,T} p(\mathcal{M}_k)}{\sum_{j=1}^{K} Z_{j,T} p(\mathcal{M}_j)}, \tag{6}$$

owing to, in general, we are not able to compute

$$Z_{k,T} = p(\mathbf{y}_{1:T}|\mathcal{M}_k) = \int_{\mathcal{X}^T} p(\mathbf{x}_{1:T}, \mathbf{y}_{1:T}|\mathcal{M}_k) d\mathbf{x}_{1:T}, \tag{7}$$

for all $k = 1, \ldots, K$. However, given an index $k \in \{1, \ldots, K\}$ and $\mathbf{y}_{1:T}$, we are able to evaluate[3]

$$p(\mathbf{x}_{1:T}, \mathbf{y}_{1:T}|\mathcal{M}_k) = \left[q_{k,1}(\mathbf{x}_1) \prod_{t=2}^{T} q_{k,t}(\mathbf{x}_t|\mathbf{x}_{t-1})\right] \left[\prod_{t=1}^{T} \ell_{k,t}(\mathbf{y}_t|\mathbf{x}_t)\right], \qquad (8)$$

so that we can also evaluate

$$\begin{aligned} p(\mathbf{x}_{1:T}|\mathbf{y}_{1:T}, \mathcal{M}_k) &= \frac{p(\mathbf{x}_{1:T}, \mathbf{y}_{1:T}|\mathcal{M}_k)}{Z_{k,T}} \\ &\propto p(\mathbf{x}_{1:T}, \mathbf{y}_{1:T}|\mathcal{M}_k), \end{aligned} \qquad (9)$$

up to a normalizing constant. On the other hand, we often cannot compute analytically integrals involving the function $p(\mathbf{x}_{1:T}|\mathbf{y}_{1:T})$. For instance, one can be interested the computation of the Minimum Mean Square Error (MMSE) estimator of the sequence of hidden states $\mathbf{x}_{1:T}$,

$$\mathbf{I}_{1:t} = E[\mathbf{x}_{1:t}] = \int_{\mathcal{X}^T} \mathbf{x}_{1:T} p(\mathbf{x}_{1:T}|\mathbf{y}_{1:T}) d\mathbf{x}_{1:T}. \qquad (10)$$

More generally, the computation of moments of $p(\mathbf{x}_{1:T}|\mathbf{y}_{1:T})$ are not analytically intractable, and its approximation is computationally demanding. In the next section, we employ Monte Carlo schemes for approximating both $p(\mathbf{x}_{1:T}|\mathbf{y}_{1:T}, \mathcal{M}_k)$ and $p(\mathcal{M}_k|\mathbf{y}_{1:T})$, for $k = 1, \ldots, K$. As a consequence, we also approximate the complete posterior pdf $p(\mathbf{x}_{1:T}|\mathbf{y}_{1:T})$ in Eq. (5).

## 3 Monte Carlo approximation via importance sampling

In order to approximate efficiently the posterior pdfs in Eqs. (3), (6) and (9) (solving the computational challenges described in the previous section), we consider the use of Monte Carlo techniques. First, we describe a *batch importance sampling* (IS) approach where, for each model $k \in \{1, \ldots, K\}$, we consider to draw $M$ possible sequences

$$\mathbf{x}_{k,1:T}^{(m)} = [\mathbf{x}_{k,1}^{(m)}, \mathbf{x}_{k,2}^{(m)}, \ldots, \mathbf{x}_{k,T}^{(m)}] \sim \varphi_k(\mathbf{x}_{1:T}),$$

from a proposal pdf $\varphi_k : \mathcal{X}^T \to \mathbb{R}$, with $m = 1, \ldots, M$. Namely, in the batch approach, we sample directly in the whole space $\mathcal{X}^T$. In the next section, we introduce the corresponding sequential scheme (working sequentially in the subspace $\mathcal{X}$). The batch IS technique is described as follows. For each index $k \in \{1, \ldots, K\}$, draw $M$ samples $\mathbf{x}_{k,1:T}^{(1)}, \ldots, \mathbf{x}_{k,1:T}^{(M)}$ from a proposal pdf $\varphi_k(\mathbf{x}_{1:T})$, where $\varphi_k : \mathcal{X}^T \to \mathbb{R}$, and assign to each sample the following importance weights

$$w_{k,T}^{(m)} = \frac{p(\mathbf{x}_{k,1:T}^{(m)}, \mathbf{y}_{1:T}|\mathcal{M}_k)}{\varphi_k(\mathbf{x}_{k,1:T}^{(m)})}, \qquad (11)$$

---

[3] For the sake of simplicity, in the following mathematical elaborations, we consider both densities $q_{k,t}$ and $\ell_{k,t}$ to be normalized w.r.t. $\mathbf{x}$ and $\mathbf{y}$ respectively, i.e., $\int_{\mathcal{X}} q_{k,t}(\mathbf{x}|\mathbf{z})d\mathbf{x} = 1$ and $\int_{\mathcal{Y}} \ell_{k,t}(\mathbf{y}|\mathbf{z})d\mathbf{y} = 1$. As a consequence, the joint pdf $p(\mathbf{x}_{1:T}, \mathbf{y}_{1:T}|\mathcal{M}_k)$ in Eq. (8) is also normalized.

with $m = 1, \ldots, M$, and $k = 1, \ldots, K$. Note that the total number of samples are $N = KM$. We can approximate

$$Z_{k,T} = p(\mathbf{y}_{1:T}|\mathcal{M}_k) = \int_{\mathcal{X}^T} p(\mathbf{x}_{k,1:T}, \mathbf{y}_{1:T}|\mathcal{M}_k) d\mathbf{x}_{k,1:T},$$

using basic IS arguments [33, 25] as[4]

$$\widehat{Z}_{k,T} = \frac{1}{M} \sum_{m=1}^{M} w_{k,T}^{(m)} \approx p(\mathbf{y}_{1:T}|\mathcal{M}_k). \tag{12}$$

We can also write $p(\mathbf{y}_{1:T}) \approx \sum_{k=1}^{K} \widehat{Z}_{k,T} p(\mathcal{M}_k)$. Furthermore, we can approximate the measure of $p(\mathbf{x}_{1:T}|\mathbf{y}_{1:T}, \mathcal{M}_k) = \frac{p(\mathbf{x}_{k,1:T}, \mathbf{y}_{1:T}|\mathcal{M}_k)}{Z_{k,T}}$ by the following particle approximation

$$\widehat{p}(\mathbf{x}_{1:T}|\mathbf{y}_{1:T}, \mathcal{M}_k) = \sum_{m=1}^{M} \bar{w}_{k,T}^{(m)} \delta(\mathbf{x}_{1:T} - \mathbf{x}_{k,1:T}^{(m)}), \tag{13}$$

where

$$\bar{w}_{k,T}^{(m)} = \frac{w_{k,T}^{(m)}}{\sum_{j=1}^{M} w_{k,T}^{(j)}} = \frac{w_{k,T}^{(m)}}{M \widehat{Z}_{k,T}}, \tag{14}$$

is the normalized weight of the $m$-th sample of the $k$-th model (normalized considering all the samples associated to the $k$-th model). Thus, given Eq. (5), we also obtain an approximation of $p(\mathbf{x}_{1:T}|\mathbf{y}_{1:T})$,

$$\widehat{p}(\mathbf{x}_{1:T}|\mathbf{y}_{1:T}) = \frac{1}{\widehat{p}(\mathbf{y}_{1:T})} \sum_{k=1}^{K} \widehat{p}(\mathbf{x}_{1:T}|\mathbf{y}_{1:T}, \mathcal{M}_k) \widehat{p}(\mathbf{y}_{1:T}|\mathcal{M}_k) p(\mathcal{M}_k)$$

$$= \frac{1}{\sum_{j=1}^{K} \widehat{Z}_{j,T} p(\mathcal{M}_j)} \sum_{k=1}^{K} \left[ \left( \frac{1}{M \widehat{Z}_{k,T}} \sum_{m=1}^{M} w_{k,T}^{(m)} \delta(\mathbf{x}_{1:T} - \mathbf{x}_{k,1:T}^{(m)}) \right) \widehat{Z}_{k,T} p(\mathcal{M}_k) \right],$$

$$= \sum_{k=1}^{K} \sum_{m=1}^{M} \left( \frac{w_{k,T}^{(m)} p(\mathcal{M}_k)}{\sum_{j=1}^{K} \sum_{i=1}^{M} w_{j,T}^{(i)} p(\mathcal{M}_j)} \right) \delta(\mathbf{x}_{1:T} - \mathbf{x}_{k,1:T}^{(m)}).$$

The last expression can be summarized as

$$\widehat{p}(\mathbf{x}_{1:T}|\mathbf{y}_{1:T}) = \sum_{k=1}^{K} \sum_{m=1}^{M} \bar{\gamma}_{k,T}^{(m)} \delta(\mathbf{x}_{1:T} - \mathbf{x}_{k,1:T}^{(m)}), \tag{15}$$

where we have denoted

$$\bar{\gamma}_{k,T}^{(m)} = \frac{w_{k,T}^{(m)} p(\mathcal{M}_k)}{\sum_{j=1}^{K} \sum_{i=1}^{M} w_{j,T}^{(i)} p(\mathcal{M}_j)}, \tag{16}$$

$$= \frac{1}{M} \frac{w_{k,T}^{(m)} p(\mathcal{M}_k)}{\sum_{j=1}^{K} \widehat{Z}_{j,T} p(\mathcal{M}_j)}. \tag{17}$$

---

[4]We consider that $\varphi_k(\mathbf{x}_{1:T})$ is normalized.

Since $\widehat{Z}_{k,T} = \frac{1}{M} \sum_{j=1}^{M} w_{k,T}^{(j)}$, it is of a particular interest from a theoretical point of view that the weight $\bar{\gamma}_{k,T}^{(m)}$ can be decomposed as[5]

$$\bar{\gamma}_{k,T}^{(m)} = \frac{w_{k,T}^{(m)}}{\sum_{j=1}^{M} w_{k,T}^{(j)}} \frac{\widehat{Z}_{k,T} p(\mathcal{M}_k)}{\sum_{j=1}^{K} \widehat{Z}_{j,T} p(\mathcal{M}_j)} = \bar{w}_{k,T}^{(m)} \bar{\rho}_{k,T}, \tag{18}$$

where $\bar{w}_{k,T}^{(m)}$ is given in Eq. (14), and $\bar{\rho}_{k,T}$ is an estimator of the posterior of the $k$-th model $p(\mathcal{M}_k|\mathbf{y}_{1:T})$ in Eq. (6),

$$\bar{\rho}_{k,T} = \frac{\widehat{Z}_{k,T} p(\mathcal{M}_k)}{\sum_{j=1}^{K} \widehat{Z}_{j,T} p(\mathcal{M}_j)} \approx p(\mathcal{M}_k|\mathbf{y}_{1:T}). \tag{19}$$

It is important to remark that the use of $K$ parallel IS schemes seems to appear naturally from the factorization $\bar{\gamma}_{k,T}^{(m)} = \bar{w}_{k,T}^{(m)} \bar{\rho}_{k,T}$. Indeed, we can also rewrite the approximation in Eq. (15) as the convex combination

$$\widehat{p}(\mathbf{x}_{1:T}|\mathbf{y}_{1:T}) = \sum_{k=1}^{K} \bar{\rho}_{k,T}\ \widehat{p}(\mathbf{x}_{1:T}|\mathbf{y}_{1:T}, \mathcal{M}_k), \tag{20}$$

where $\bar{\rho}_{k,T}$ is the normalized weight of the $k$-th model. Finally, for instance, the computation of the MMSE estimator $\widehat{\mathbf{x}}_{1:T}$ in Eq. (10) is approximated as

$$\mathbf{I}_{1:T} \approx \widehat{\mathbf{I}}_{1:T} = \sum_{k=1}^{K} \sum_{m=1}^{M} \bar{\gamma}_{k,T}^{(m)} \mathbf{x}_{k,1:T}^{(m)}, \tag{21}$$

or with the equivalent two-stage formula

$$\begin{cases} \widetilde{\mathbf{I}}_{k,1:T} &= \sum_{m=1}^{M} \bar{w}_{k,T}^{(m)} \mathbf{x}_{k,1:T}^{(m)}, \\ \widehat{\mathbf{I}}_{1:T} &= \sum_{k=1}^{K} \bar{\rho}_{k,T} \widetilde{\mathbf{I}}_{k,1:T}, \end{cases} \tag{22}$$

where $\widetilde{\mathbf{I}}_{k,1:T}$ represents the approximated MMSE estimator considering only the $k$-th model. The IS procedure above can easily reformulated within a sequential framework, as we describe in the next section.

## 3.1 Sequential Importance Sampling (SIS) and marginal likelihood estimation

The IS method can be alternatively performed in a sequential manner, i.e., providing an approximation of the posterior pdf at each iteration $t$ using the previous approximation at the iteration $t-1$. Let us consider an index $k \in \{1, \ldots, K\}$. Observing the following recursive relationship between the posterior pdfs at $t-1$ and $t$ [13],

$$p(\mathbf{x}_{1:t}|\mathbf{y}_{1:t}, \mathcal{M}_k) = \frac{\ell_{k,t}(\mathbf{y}_t|\mathbf{x}_t) q_{k,t}(\mathbf{x}_t|\mathbf{x}_{t-1})}{p(\mathbf{y}_t|\mathbf{y}_{1:t-1}, \mathcal{M}_k)} p(\mathbf{x}_{1:t-1}|\mathbf{y}_{1:t-1}, \mathcal{M}_k), \tag{23}$$

---
[5]Note also that $\sum_{k=1}^{K} \sum_{m=1}^{M} \bar{\gamma}_{k,T}^{(m)} = 1$.

(see A for further details), we can build the empirical approximation $\widehat{p}(\mathbf{x}_{1:t}|\mathbf{y}_{1:t},\mathcal{M}_k)$ as

$$\widehat{p}(\mathbf{x}_{1:t}|\mathbf{y}_{1:t},\mathcal{M}_k) = \frac{1}{M\widehat{Z}_{k,t}} \sum_{i=1}^{M} w_{k,t}^{(i)} \delta(\mathbf{x}_{1:t} - \mathbf{x}_{k,1:t}^{(i)}), \qquad (24)$$

given the previous $\widehat{p}(\mathbf{x}_{1:t-1}|\mathbf{y}_{1:t-1},\mathcal{M}_k)$. Recall that, we also obtain an estimator of $p(\mathbf{y}_{1:t}|\mathcal{M}_k)$, as $\widehat{Z}_{k,t} = \frac{1}{M}\sum_{m=1}^{M} w_{k,t}^{(m)}$. Let us consider a proposal density $\varphi_k : \mathcal{X}^T \to \mathbb{R}$ factorizes as

$$\varphi_k(\mathbf{x}_{k,1:T}) = \phi_{k,1}(\mathbf{x}_{k,1}) \prod_{t=2}^{T} \phi_{k,t}(\mathbf{x}_{k,t}|\mathbf{x}_{k,t-1}), \qquad (25)$$

with $\phi_{k,t} : \mathcal{X} \to \mathbb{R}$ for $t = 1,\ldots,T$. Given Eq. (23), we can infer a recursive relationship between two importance weights at consecutive iterations [13],

$$\begin{aligned}
w_{k,t}^{(m)} &= w_{k,t-1}^{(m)} \frac{\ell_{k,t}(\mathbf{y}_t|\mathbf{x}_{k,t}^{(m)}) q_{k,t}(\mathbf{x}_{k,t}^{(m)}|\mathbf{x}_{k,t-1}^{(m)})}{\phi_{k,t}(\mathbf{x}_{k,t}^{(m)}|\mathbf{x}_{k,1:t-1}^{(m)})}, \\
&= w_{k,t-1}^{(m)} \lambda_{k,t}^{(m)} = \prod_{\tau=1}^{t} \lambda_{k,\tau}^{(m)},
\end{aligned} \qquad (26)$$

where

$$\lambda_{k,t}^{(m)} = \frac{\ell_{k,t}(\mathbf{y}_t|\mathbf{x}_{k,t}^{(m)}) q_{k,t}(\mathbf{x}_{k,t}^{(m)}|\mathbf{x}_{k,t-1}^{(m)})}{\phi_{k,t}(\mathbf{x}_{k,t}^{(m)}|\mathbf{x}_{k,1:t-1}^{(m)})},$$

and $\mathbf{x}_{k,t}^{(m)} \sim \phi_{k,t}(\mathbf{x}_{k,t}^{(m)}|\mathbf{x}_{k,1:t-1}^{(m)})$, and $m = 1,\ldots,M$. Therefore, given the recursive expression of $w_{k,t}^{(m)}$, we can rewrite the estimator $\widehat{Z}_{k,t}$ as

$$\widehat{Z}_{k,t} = \frac{1}{M} \sum_{m=1}^{M} w_{k,t}^{(m)} = \frac{1}{M} \sum_{m=1}^{M} \left[ \prod_{\tau=1}^{t} \lambda_{k,\tau}^{(m)} \right]. \qquad (27)$$

However, the estimator of $p(\mathbf{y}_{1:t}|\mathcal{M}_k)$ above has not a unique formulation. Indeed, it is possible to obtain an approximation of the denominator in Eq. (23) (see B),

$$\widehat{p}(\mathbf{y}_t|\mathbf{y}_{1:t-1},\mathcal{M}_k) = \sum_{i=1}^{M} \bar{w}_{k,t-1}^{(i)} \lambda_{k,t}^{(i)},$$

where $\bar{w}_{k,t-1}^{(i)} = \frac{w_{k,t-1}^{(i)}}{\sum_{n=1}^{N} w_{k,t-1}^{(n)}}$ are the normalized weights at $t-1$-th iteration. As a consequence, since $p(\mathbf{y}_{1:t}|\mathcal{M}_k) = \prod_{\tau=1}^{t} p(\mathbf{y}_\tau|\mathbf{y}_{1:\tau-1},\mathcal{M}_k)$, we have a second possible formulation of the estimator of $p(\mathbf{y}_{1:t}|\mathcal{M}_k)$,

$$\widetilde{Z}_{k,t} = \prod_{\tau=1}^{t} \left[ \sum_{j=1}^{M} \bar{w}_{k,\tau-1}^{(j)} \lambda_{k,\tau}^{(j)} \right] = \prod_{\tau=1}^{t} \left[ \frac{\sum_{j=1}^{M} w_{k,\tau}^{(j)}}{\sum_{j=1}^{M} w_{k,\tau-1}^{(j)}} \right]. \qquad (28)$$

In B, we show a complete derivation of the estimator $\widetilde{Z}_{k,t}$ and that $\widetilde{Z}_{k,t} \equiv \widehat{Z}_{k,t}$, as one could easily realize from Eq. (28).

**Remark 1.** *Note that, in SIS, there are two possible equivalent formulation of the estimator of* $p(\mathbf{y}_{1:t}|\mathcal{M}_k)$, *i.e.,* $\widehat{Z}_{k,t}$ *in Eqs. (27)* $\widetilde{Z}_{k,t}$ *in Eq. (28), and they are equivalent,* $\widetilde{Z}_{k,t} \equiv \widehat{Z}_{k,t}$.

In any case, since $\widehat{p}(\mathbf{y}_{1:t}) = \sum_{j=1}^{M} Z_{j,t} p(\mathcal{M}_j)$, the model weight $\bar{\rho}_{k,t} = \frac{\widehat{p}(\mathbf{y}_{1:t}, \mathcal{M}_k)}{\widehat{p}(\mathbf{y}_{1:t})}$ employed in Eqs (19)-(20) can be expressed as

$$\bar{\rho}_{k,t} = \frac{\widehat{Z}_{k,t} p(\mathcal{M}_k)}{\sum_{j=1}^{M} \widehat{Z}_{j,t} p(\mathcal{M}_j)} \approx p(\mathcal{M}_k | \mathbf{y}_{1:t}). \quad (29)$$

The weights $\bar{\rho}_{k,t}$ above are then used for computing the global estimator at the $t$-th iteration, similarly in Eq. (22).

## 3.2 Sequential Importance Resampling (SIR) and marginal likelihood estimation

In several algorithms, such as the *sequential Monte Carlo* (SMC) methods, resampling steps are performed within SIS schemes for avoiding the degeneracy of the weights [13, 15]. Let us denote as $\bar{\mathbf{x}}_{k,1:t}^{(j)} \in \{\mathbf{x}_{k,1:t}^{(1)}, \ldots, \mathbf{x}_{k,1:t}^{(M)}\}$, a resampled particle at the iteration $t$ (resampled according to the normalized weights $\bar{w}_{k,t-1}^{(j)}$, $j = 1, \ldots, M$, at the $t$-th iteration). The unnormalized importance weights of the resampled particles, denoted as $\alpha_{k,t}^{(j)}$, $j = 1, \ldots, M$, are set to the same value [13, 15, 28], i.e,[6]

$$\alpha_{k,t}^{(1)} = \alpha_{k,t}^{(2)} = \ldots = \alpha_{k,t}^{(M)}.$$

A proper value for the (unnormalized) importance weight $\alpha_{1:t-1}^{(j)}$ associated with the $j$-th resampled particle [28] is

$$\alpha_{k,t}^{(j)} = \widehat{Z}_{k,t} = \frac{1}{N} \sum_{i=1}^{N} w_{k,t}^{(i)}, \quad \forall j = 1, \ldots, M. \quad (30)$$

One reason why this seems a suitable choice, for instance, is that defining the following weights

$$\xi_{k,t}^{(m)} = \begin{cases} w_{k,t}^{(m)}, & \text{without resampling at } t\text{-th iteration,} \\ \alpha_{k,t}^{(m)}, & \text{with resampling at } t\text{-th iteration.} \end{cases} \quad (31)$$

then, in any case, $\frac{1}{N} \sum_{n=1}^{N} \xi_{k,t}^{(j)} = \widehat{Z}_{k,t}$, as expected. In general, the resampling steps are not applied at each iterations, but only when some statistical criterion is satisfied [8, 13, 15] (e.g., see Section 4). The recursive expression of the weights for SIR becomes

$$\xi_{k,t}^{(m)} = \xi_{k,t-1}^{(m)} \lambda_{k,t}^{(m)}, \quad (32)$$

where

$$\xi_{k,t-1}^{(m)} = \begin{cases} \xi_{k,t-1}^{(m)}, & \text{without res. at } (t-1)\text{-th iter.,} \\ \widehat{Z}_{k,t-1}, & \text{with res. at } (t-1)\text{-th iter.} \end{cases} \quad (33)$$

---
[6]This is a proper choice, but it is not unique; see *The concept of weighted samples* in [25, Chapter 2] or [33, Section 14.2].

Table 1: **Main notation for MAPF.**

| | |
|---|---|
| $K$ | Number of parallel PFs (and different models). |
| $T$ | Total number of iterations of each PF. |
| $N$ | Total a number of particles distributed among the PFs. |
| $M_{k,t}$ | Number of particles of the $k$-th PF at the iteration $t$. |
| $T_V$ | Refreshing parameter. |
| $\widetilde{Z}_{k,t}, \widehat{Z}_{k,t}$ | estimator of $p(\mathbf{y}_{1:t}|\mathcal{M}_k)$ (two equivalent formulations). |
| $w_{k,t}^{(m)}$ | Importance weight assigned to the sample $\mathbf{x}_{k,t}^{(m)}$. |
| $\bar{w}_{k,t}^{(m)}$ | Weight assigned to the sample $\mathbf{x}_{k,t}^{(m)}$, normalized w.r.t. the $M_{k,t}$ weights of $k$-th filter. |
| $\bar{\rho}_{k,t}$ | Approximation of $p(\mathcal{M}_k|\mathbf{y}_{1:t})$. |
| $\widetilde{\rho}_{k,t}$ | Approximation of $\widetilde{p}(\mathcal{M}_k|\mathbf{y}_{1:t})$, considering an updated prior $\widetilde{p}(\mathcal{M}_k)$. |
| $\bar{\gamma}_{k,t}^{(m)}$ | Global normalized weight ($\bar{\gamma}_{k,t}^{(m)} = \bar{w}_{k,t}^{(m)}\bar{\rho}_{k,t}$ or $\bar{\gamma}_{k,t}^{(m)} = \bar{w}_{k,t}^{(m)}\widetilde{\rho}_{k,t}$). |
| $\mathbf{I}_{1:t} = E[\mathbf{x}_{1:t}]$ | MMSE estimator of the hidden state $\mathbf{x}_{1:t}$. |
| $\widetilde{\mathbf{I}}_{1:t}$ | Partial approximation of $\mathbf{I}_{1:t}$ of the $k$-th PF. |
| $\widehat{\mathbf{I}}_{1:t}$ | Global approximation of $\mathbf{I}_t$. |

**Remark 2.** *With the recursive definition of the weights $\xi_{k,t}^{(m)}$'s in Eqs. (32)-(33), the two possible estimators of the marginal likelihood $p(\mathbf{y}_{1:t}|\mathcal{M}_k)$ are the same within SIR, as well. The estimators are*

$$\widehat{Z}_{k,t} = \frac{1}{M}\sum_{m=1}^{M} \xi_{k,t-1}^{(m)}\lambda_{k,t}^{(m)}, \quad \widetilde{Z}_{k,t} = \prod_{\tau=1}^{t}\left[\sum_{j=1}^{M}\bar{\xi}_{k,\tau-1}^{(j)}\lambda_{k,\tau}^{(j)}\right] \quad (34)$$

*where $\bar{\xi}_{k,t-1}^{(j)} = \frac{\xi_{k,t-1}^{(j)}}{\sum_{m=1}^{M}\xi_{k,t-1}^{(m)}}$. They are equivalent and valid estimators [28] as shown in B.*

So far, in this section, we have considered a specific value of the index $k \in \{1,\ldots,K\}$. However, for our purpose, we need of all the model weights $\bar{\rho}_{k,t}$, for all values $k = 1,\ldots,K$. Thus, a practical implementation employing $K$ parallel particle filters appears natural. In the next section, we describe the proposed scheme in details.

## 4 Model Averaging Particle Filters

The factorization of the global normalized weights $\bar{\gamma}_{k,t}^{(m)} = \bar{w}_{k,t}^{(m)}\bar{\rho}_{k,t}$ suggests the use of $K$ parallel IS schemes. Indeed, each IS scheme can compute independently $\bar{w}_{k,t}^{(m)}$ and $\widehat{Z}_{k,t}$, and then they merge all the information for calculating $\bar{\rho}_{k,t}$ (see Figure 1). Therefore, we consider $K$ parallel particle filters (as in [8, 12, 13], for instance) using the transition model as proposal pdf[7]

$$\phi_{k,t}(\mathbf{x}_{k,t}|\mathbf{x}_{k,1:t-1}) = q_{k,t}(\mathbf{x}_{k,t}|\mathbf{x}_{k,t-1}),$$

each one tracking a different states-space model $\mathcal{M}_k$, for $k = 1,\ldots,K$. The $K$ parallel PFs cooperate for providing a unique global approximation of the complete posterior, as described in the previous sections. As a consequence, an approximation $\bar{\rho}_{k,T}$ of the model posterior densities $p(\mathcal{M}_k|\mathbf{y}_{1:T})$ is also provided.

---
[7]More sophisticated PF techniques could be also employed, each one addressing a different model.

Table 1 summarizes the main notation used in the *Model Averaging Particle Filter* (MAPF) and Table 2 provides an exhaustive description of MAPF. The computational effort is distributed adaptively among the parallel PFs, proportionally to the (approximated) posterior pdf of the model $\mathcal{M}_k$ given the current set of data $\mathbf{y}_{1:t}$, i.e., $\bar{\rho}_{k,t} \approx p(\mathcal{M}_k|\mathbf{y}_{1:t})$. More specifically, when a resampling step is performed, the number of particles of each filter is adapted online, taking into account an approximation of the probability $p(\mathcal{M}_k|\mathbf{y}_{1:t})$. However, the overall computational cost remains invariant (pre-established by the user in advance) since total number of particles denote as $N$ does not vary, i.e.,

$$\sum_{j=1}^{K} M_{j,t} = M_{1,t} + M_{2,t} + \ldots + M_{K,t} = N, \qquad \text{for} \quad t = 1, \ldots, T. \tag{35}$$

Since $M_{k,t+1} = \lfloor \bar{\rho}_{k,t} \rfloor$, when a resampling step is applied, (or $M_{k,t+1} = M_{k,t}$, without) in general we have $N' = N - \sum_{j=1}^{K} M_{j,t+1} \geq 0$. When $N' > 0$, the remaining $N'$ particles can be assigned to the filters in different ways: for instance, with probability $\bar{\rho}_{k,t}$, i.e., set $M_{k,t+1} = M_{k,t+1} + N'$ so that $\sum_{j=1}^{K} M_{j,t+1} = N$. Alternatively, the use of a minimum number of particles for each filter can be considered (avoiding, in this way, the stop of some filter with no particles assigned).[8] The MMSE estimator $\widehat{\mathbf{x}}_t$ in Eq. (10) can be approximated using the Eq. (21) or (22), i.e.,

$$\widehat{\mathbf{I}}_t = \sum_{k=1}^{K} \bar{\rho}_{k,t} \widetilde{\mathbf{I}}_{k,t} \quad \text{where} \quad \widetilde{\mathbf{I}}_{k,t} = \sum_{i_k=1}^{M_{k,t}} \bar{w}_{k,t}^{(i_k)} \mathbf{x}_{k,t}^{(i_k)}, \tag{36}$$

where $\widetilde{\mathbf{I}}_{k,t}$ is provided by each filter independently, as shown in Figure 1. We have denoted as $\bar{\mathbf{x}}_{k,1:t}^{(i_k)} \in \{\mathbf{x}_{k,1:t}^{(1)}, \ldots, \mathbf{x}_{k,1:t}^{(M_{k,t})}\}$ the samples after a resampling step. A resampling step is applied at each iteration that the Effective Sample Size (ESS) is smaller than a threshold value ($\epsilon N$ where $|\epsilon| \leq 1$). We adapt some approximations of ESS proposed in literature [13, 20, 27, 26] to the MAPF context, hence for instance the possible conditions are

$$\widehat{ESS}(\bar{\gamma}) = \frac{1}{\sum_{k=1}^{K} \sum_{m=1}^{M_{k,t}} \left(\bar{\gamma}_{k,t}^{(m)}\right)^2} \leq \epsilon N, \quad \text{or} \quad \widehat{ESS}(\bar{\gamma}) = \frac{1}{\max_{\forall m,k} \bar{\gamma}_{k,t}^{(m)}} \leq \epsilon N \tag{37}$$

where $\bar{\gamma}_{k,t}^{(m)}$ are the global weights in Eq. (16). Namely, when $\widehat{ESS} \leq \epsilon N$, a resampling step is performed.

**Remark 3.** *Note that ESS approximations in Eq. (37), $\widehat{ESS}(\bar{\gamma})$, involve the use of the global weights, $\bar{\gamma}_{k,t}^{(m)}$, so that the decision of applying resampling in a specific iteration is also a cooperative task carried out by the parallel filters.*

---

[8]Observe that for the algorithm in Table 2, it is necessary that $M_{k,t} \geq 2$, $\forall k, t$. In the equivalent MAPF formulation given in C, this constraint is not required.

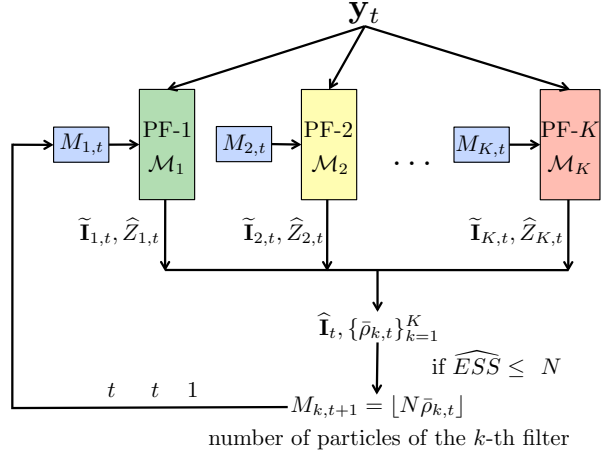

Figure 1: Graphical overview of the MAPF scheme in Table 2 for approximating the MMSE estimator $\widehat{\mathbf{I}}_t$ and the posterior of each model given the observed data. Note that both $\widehat{ESS}$ formulas in Eq. (37) take into account the (normalized) weights $\bar{\gamma}_{k,t}^{(i_k)} = \bar{w}_{k,t}^{(i_k)} \bar{\rho}_{k,t}$, for all $i_k = 1, ..., M_{k,t}$ and $k$.

Figure 2 depicts another graphical representation of the MAPF algorithm in Table 2 with $K = 2$ filters, when a resampling step is employed. The filters interact through the adaptation of numbers of particles of each filter, $M_{k,t}$ (and also providing jointly the global MMSE estimator). However, Figure 2 shows that the resampling steps are performed independently by each filter (in the figure $K = 2$), i.e., no particles are exchanged among the filters. Further considerations about MAPF are provided in C.

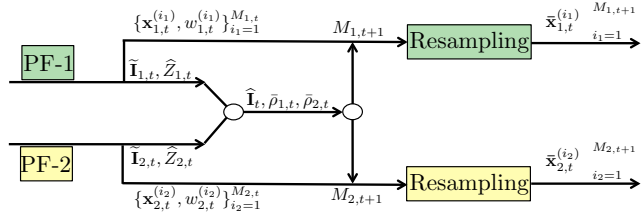

Figure 2: Another graphical representation of the MAPF scheme, with $K = 2$ parallel particle filters. A fusion center returns, $\widehat{\mathbf{I}}_t$, $\bar{\rho}_{k,t}$, and the numbers of particles $M_{k,t+1}$, $k = 1, 2$, of the next iterations. Then, the resampling steps are performed separately.

**Remark 4.** *The step 4 of MAPF in Table 2 can be interpreted as a double resampling: a first resampling considering the model weights, $\bar{\rho}_{k,t}$, adapting the number of particles and the second one considering the normalized weights, $\bar{w}_{k,t}^{(m)}$ within a filter. However, in this scheme the exchange of particles among the filters is not allowed. In Section 5 another resampling procedure is considered, using directly the global weights $\bar{\gamma}_{k,t}^{(m)}$, where the exchange of particles is possible.*

The remark above is strictly connected to the considerations in C.

Table 2: **Model Averaging Particle Filters (MAPF)**.

---

- **Initialization:** Set $M_{k,1} = \frac{N}{K} \geq 2$, $\xi_{k,0}^{(i_k)} = 1$, and choose the initial states $\bar{\mathbf{x}}_{k,0}^{(i_k)}$, for all $i_k = 1, \ldots, M_{k,1}$ and $k = 1, \ldots, K$.
- **For** $t = 1, \ldots, T$:

  1. **Propagation:** Draw $\mathbf{x}_{k,t}^{(i_k)} \sim q_{k,t}(\mathbf{x}|\bar{\mathbf{x}}_{k,t-1}^{(i_k)})$, for $i_k = 1, \ldots, M_{k,t}$ and $k = 1, \ldots, K$.

  2. **Particle Weighting:** Compute the weights and normalized them,

  $$w_{k,t}^{(i_k)} = \xi_{k,t-1}^{(i_k)} \ell_{k,t}(\mathbf{y}_t|\mathbf{x}_{k,t}^{(i_k)}), \text{ and } \bar{w}_{k,t}^{(i_k)} = \frac{w_{k,t}^{(i_k)}}{\sum_{j=1}^{M_{k,t}} w_{k,t}^{(j)}}, \qquad (38)$$

  for $i_k = 1, \ldots, M_{k,t}$, $k = 1, \ldots, K$ and $\xi_{k,t-1}^{(i_k)}$ is defined in Eqs. (32)-(33).

  3. **Model Weighting:** Compute, for $k = 1, \ldots, K$,

  $$\widehat{Z}_{k,t} = \frac{1}{M_{k,\tau}} \sum_{j=1}^{M_{k,\tau}} w_{k,t}^{(j)}, \text{ and } \bar{\rho}_{k,t} = \frac{\widehat{Z}_{k,t} p(\mathcal{M}_k)}{\sum_{j=1}^{K} \widehat{Z}_{j,t} p(\mathcal{M}_j)}. \qquad (39)$$

  Alternatively, the estimator $\widehat{Z}_{k,t}$ in Eq. (34) can be used.

  4. **Adaptation and Resampling:** For each filter, $k = 1, \ldots, K$:
     - If the condition (37) is fulfilled,
       (a) Set $M_{k,t+1} = \lfloor N\bar{\rho}_{k,t} \rfloor$, and distribute the remaining $N' = N - \sum_{j=1}^{K} M_{j,t+1}$ particles among the $K$ filters according to some pre-established criterion (see Section 4).
       (b) Draw $M_{k,t+1}$ samples, $\bar{\mathbf{x}}_{k,1:t}^{(1)}, \ldots, \bar{\mathbf{x}}_{k,1:t}^{(M_{k,t+1})}$, from

  $$\widehat{p}(\mathbf{x}_{1:t}|\mathbf{y}_{1:t}, \mathcal{M}_k) = \sum_{i_k=1}^{M_{k,t}} \bar{w}_{k,t}^{(i_k)} \delta(\mathbf{x}_{1:t} - \mathbf{x}_{k,1:t}^{(i_k)}).$$

     - Otherwise, set $M_{k,t+1} = M_{k,t}$ and $\bar{\mathbf{x}}_{k,1:t}^{(i_k)} = \mathbf{x}_{k,1:t}^{(i_k)}$, for all $i_k$.

  5. **Output:** Return $\{\mathbf{x}_{k,1:t}^{(i_k)}, w_{k,t}^{(i_k)}, \bar{\rho}_{k,t}, \bar{\gamma}_{k,t}^{(i_k)}\}$, $i_k = 1, \ldots, M_{k,t}$ and $k = 1, \ldots, K$.

---

## 4.1 Factorization of the global weights in MAPF: Empirical Bayes approach

Recalling the definition in Eq. (16),

$$\bar{\gamma}_{k,t}^{(m)} = \frac{w_{k,t}^{(m)} p(\mathcal{M}_k)}{\sum_{j=1}^{K} \sum_{i=1}^{M} w_{j,t}^{(i)} p(\mathcal{M}_j)},$$

note that in this case we have (due to the use of different number of particles in the parallel filters)

$$\bar{\gamma}_{k,t}^{(m)} = \frac{w_{k,t}^{(m)} p(\mathcal{M}_k)}{\sum_{j=1}^{K} M_{j,t} \widehat{Z}_{j,t} p(\mathcal{M}_j)} = \frac{w_{k,t}^{(m)}}{\sum_{j=1}^{M_{k,t}} w_{k,t}^{(j)}} \frac{M_{k,t} \widehat{Z}_{k,t} p(\mathcal{M}_k)}{\sum_{j=1}^{K} M_{j,t} \widehat{Z}_{j,t} p(\mathcal{M}_j)} = \bar{w}_{k,t}^{(m)} \widetilde{\rho}_{k,t}, \qquad (40)$$

where

$$\widetilde{\rho}_{k,t} = \frac{\widehat{Z}_{k,t} M_{k,t} p(\mathcal{M}_k)}{\sum_{j=1}^{K} \widehat{Z}_{j,t} M_{j,t} p(\mathcal{M}_j)} = \frac{\widehat{Z}_{k,t} \widetilde{p}(\mathcal{M}_k)}{\sum_{j=1}^{K} \widehat{Z}_{j,t} \widetilde{p}(\mathcal{M}_j)}, \quad (41)$$

and $\widetilde{p}(\mathcal{M}_j) = M_{j,t} p(\mathcal{M}_j)$ plays the role of an updated prior pdf as in an empirical Bayes approach [6].

## 4.2 Sub-cases of interest

It is interesting to remark that, if the likelihood function is common to all the filters, i.e., $\ell_{k,t}(\mathbf{y}_t|\mathbf{x}_t) = \ell_t(\mathbf{y}_t|\mathbf{x}_t)$ for all $k$, then MAPF can be seen as a PF with adaptive prior pdf. Thus, the complete likelihood is $p(\mathbf{y}_{1:t}|\mathbf{x}_{1:t}) = \prod_{\tau=1}^{t} \ell_\tau(\mathbf{y}_\tau|\mathbf{x}_\tau)$, and the complete prior pdf of $k$-th model, is[9]

$$p(\mathbf{x}_{1:t}|\mathcal{M}_k) = q_{k,1}(\mathbf{x}_1) \prod_{\tau=2}^{t} q_{k,\tau}(\mathbf{x}_\tau|\mathbf{x}_{\tau-1}). \quad (42)$$

Namely, in this setup, the model selection problem becomes as the problem of a suitable choice of a prior pdf for our model. This specific case is particularly interesting from a practical point of view, as we discuss below. Since in this case $p(\mathbf{x}_{1:t}|\mathbf{y}_{1:t},\mathcal{M}_k) = p(\mathbf{y}_{1:t}|\mathbf{x}_{1:t}) p(\mathbf{x}_{1:t}|\mathcal{M}_k)$, the complete posterior $p(\mathbf{x}_{1:t}|\mathbf{y}_{1:t})$ in Eq. (4) can be rewritten as

$$p(\mathbf{x}_{1:t}|\mathbf{y}_{1:t}) = p(\mathbf{y}_{1:t}|\mathbf{x}_{1:t}) \left[ \sum_{k=1}^{K} \zeta_k\, p(\mathbf{x}_{1:t}|\mathcal{M}_k) \right], \quad (43)$$

where we have denoted $\zeta_k = p(\mathcal{M}_k|\mathbf{y}_{1:T})$. Noting that $\sum_{k=1}^{K} \zeta_k = 1$, we can interpret this framework as using a unique model with the dynamic-prior pdf defined by the following mixture, $p(\mathbf{x}_{1:t}) = \sum_{k=1}^{K} \zeta_k\, p(\mathbf{x}_{1:t}|\mathcal{M}_k)$. In this case, when the resampling is applied, the MAPF approach can also be interpreted as a *unique* particle filter with an *adaptive* prior pdf approximating the prior mixture $\sum_{k=1}^{K} \zeta_k\, p(\mathbf{x}_{1:t}|\mathcal{M}_k)$, using the approximated weights $\bar{\rho}_{k,t} \approx \zeta_k$. Furthermore, if all the models are the same equal to the true one $\mathcal{T}$, i.e., $\mathcal{M}_1 = \mathcal{M}_2 = \ldots = \mathcal{M}_K = \mathcal{T}$, then MAPF described a distributed PF scheme [8, 12, 15] which $K$ parallel PFs cooperate for providing a global estimator, $\widehat{\mathbf{I}}_{1:t} = \sum_{k=1}^{K} \bar{\rho}_{k,t} \widetilde{\mathbf{I}}_{k,1:t}$. The computational effort is distributed in order to foster the filters that are providing the best performance, in the specific run.

## 5 MAPF for time-varying models

The MAPF scheme can be easily modified for applying it in a time-varying model setting, where at some unknown iterations $t_1^* \leq t_2^* \leq t_3^* \ldots$ the true model $\mathcal{T}$ generating $\mathbf{x}$'s and $\mathbf{y}$'s changes, i.e., we have $\mathcal{T}_1 \to \mathcal{T}_2 \to \mathcal{T}_3 \ldots$ etc. For example, a traveller may switch from walking to riding the bus, and both the dynamics and observations will vary accordingly. The change detection problem [3] can be also considered as an interesting particular case. For instance, see the numerical example in Section 6.1.

---

[9]Note that, in a Bayesian setting, the dynamic models play the role of prior pdf.

The simplest way for handling this scenario is to consider a window of $T_V \geq 1$ iterations for computing the values $\widehat{Z}_{k,t}$, for $k = 1, \ldots, K$. In this case we modify the computation of $\widehat{Z}_{k,t}$: the simplest possibility is to refresh all the $\widehat{Z}_{k,t}$'s each $T_V$ iterations considering only the incremental weights (as forcing $\xi_{k,t-1} = \widehat{Z}_{k,t-1} = 1$)[10], i.e.,

$$\widehat{Z}_{k,t^*} = \frac{1}{M_{k,t^*}} \sum_{m=1}^{M_{k,t^*}} \lambda_{k,t^*}^{(m)}, \quad t^* = rT_V, \quad r \in \mathbb{N}. \tag{44}$$

On the one hand, with a small $T_V$, the algorithm is able to detect quickly the change in the model $\mathcal{T}$, although the approximation of the posteriors of the model given the data become more unstable (since the posterior takes into account a smaller number of observations). On the other hand, with a bigger $T_V$, the algorithm is more stable but the detection of model updates is slower. In this scenario, every $T_V$ iterations, the number of particles of each filter should be refreshed, setting $M_{k,t} = \frac{N}{K}$ for all $k = 1, \ldots, K$. Moreover, in order to recover lost filters, a joint resampling should be performed, considering the weights $\bar{\gamma}_{k,t}^{(i_k)} = \bar{w}_{k,t}^{(i_k)} \widetilde{\rho}_{k,t}$, $i_k = 1, \ldots, M_{k,t}$ and $k = 1, \ldots, K$ (instead of considering only $\bar{w}_{k,t}^{(i_k)}$; see Remark 4). This allows the exchange of particles among the different filters. Namely, one could draw $N$ particles from the global approximation at the $t$-th iteration in Eq. (15),

$$\widehat{p}(\mathbf{x}_{1:t}|\mathbf{y}_{1:t}) = \sum_{k=1}^{K} \sum_{i_k=1}^{M_{k,t}} \bar{\gamma}_{k,t}^{(i_k)} \delta(\mathbf{x}_{1:t} - \mathbf{x}_{k,1:t}^{(i_k)}), \tag{45}$$

where $\bar{\gamma}_{k,t}^{(i_k)}$ are in Eqs. (16)-(18). Figure 3 summarizes the suggested approach for handling the selection of time-varying models. The application of this refreshing strategy could be useful also in the standard model selection setting, without changes in the true model: it avoids numerical problems and can increase the robustness of the MAPF technique.

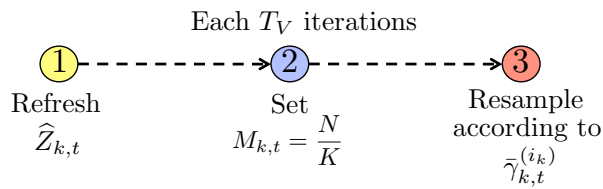

Figure 3: Refreshing strategy for time-varying model settings.

Namely, $M_{k,t} = \frac{N}{K}$ particles, for each filters, are obtained sampling $\frac{N}{K}$ times from $\widehat{p}(\mathbf{x}_{1:t}|\mathbf{y}_{1:t})$ in Eq. (45). Finally, let us consider for simplicity the sub-case of common likelihood function. The improvement in the performance provided by MAPF with respect to a standard filter can be considerable in tracking applications. This is owing to, for instance, riding a bus hugely constrains both the dynamic and observation model. Indeed, simply knowing the bus routes beforehand, the movement can be restricted by a whole dimension, since the trajectory of the bus is fixed. The mathematical descriptions of different modalities of mobility can be calibrated in advance from publicly available data. In the numerical simulations in Section 6.3, for example, we use shows the real-time position of buses in the city of Helsinki (http://live.mattersoft.fi/hsl/).

---

[10]Alternatively a sliding window of iterations can be considered, i.e., $\widehat{Z}_{k,t} \equiv \widetilde{Z}_{k,t} = \prod_{\tau=t-T_V+1}^{t} \left[ \sum_{j=1}^{M_{k,\tau}} \bar{\xi}_{k,\tau-1}^{(j)} \lambda_{k,\tau}^{(j)} \right]$.

**Automatic adaptive selection of the refreshing steps.** The parameter $T_V$ can also be adaptively and automatically tuned. Indeed, an automatic approach can be used in order to decide when applying the refreshing steps. For instance, we can consider to apply the refreshing procedure, summarized in Fig. 3 (including the global resampling that involves $\bar{\gamma}_{k,t}^{(m)}$'s), as alternative of the standard resampling steps (that employ $\bar{w}_{k,t}^{(m)}$'s). More specifically, when one of the chosen ESS condition in (37) is fulfilled, i.e., $\widehat{ESS}(\bar{\gamma}) \leq \epsilon N$, with probability $p_r$ we apply the refreshing strategy (with the global resampling) otherwise, with probability $1 - p_r$, we apply a standard resampling within each filter.[11] Figure 4 summarizes this procedure. Finally, note also that a mix strategy can be also employed: the adaptive procedure can be applied and the refreshing can be forced at some specific iterations (in order to increase the robustness of the method).

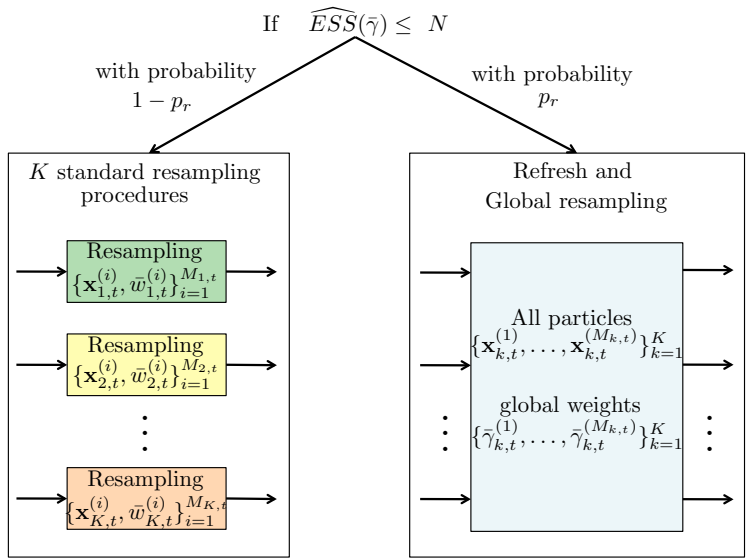

Figure 4: Graphical representation of the automatic and adaptive procedure in order to select the refreshing steps.

## 6 Numerical simulations

We consider three different numerical experiments in order to test the MAPF algorithm. The first two numerical examples employ artificial data whereas the last one involves real data.[12] More specifically, in the first example we study the ability of MAPF in a time-varying model scenario. The second example considers the parameter selection and tracking problem in a state-space model (the parameter selection can be interpreted as a sub-case of the model selection problem). The last numerical experiment applies MAPF in a urban mobility context, considering real-world data involving smartphone-based tracking.

---

[11]Our experience suggests the choice $0.05 \leq p_r \leq 0.2$.

[12] A preliminary MATLAB code of MAPF related to the first two numerical examples is provided at https://www.mathworks.com/matlabcentral/fileexchange/58597-model-averanging-particle-filter.

## 6.1 Online model selection for time series

As a first example, we consider an inference problem given two possible systems of stochastic equations for modeling a time series ($x_t \in \mathbb{R}$, $t \in \mathbb{N}$). The goal is to estimate sequentially the hidden sequence $x_{1:T}$ and also to recognize, in an online fashion, which model generates the received measurements $y_{1:T}$, between

$$\mathcal{M}_1 : \begin{cases} x_{1,t} = \frac{ax_{1,t}}{1+bx_{1,t}^2} + v_{1,t}, \\ y_t = x_{1,t} + u_{1,t}, \end{cases} \quad (46)$$

$$\mathcal{M}_2 : \begin{cases} x_{2,t} = x_{2,t-1} + v_{2,t}, \\ y_t = \exp\{-c\, x_{2,t}\} + u_{2,t}, \end{cases} \quad (47)$$

with, $a = -10$, $b = 3$, $c = 0.2$ and $t = 1, \ldots, T = 500$. The variables $v_{k,t} \sim \mathcal{N}(0,1)$ and $u_{k,t} \sim \mathcal{N}(0, \frac{1}{2})$, $k = 1, 2$ represent Gaussian perturbations.

We set that the first 250 observations are generated from $\mathcal{M}_1$ and the remaining from $\mathcal{M}_2$. Namely, for $t \leq 250$ the true model is $\mathcal{T}_{t \leq 250} \equiv \mathcal{M}_1$ whereas, for $t > 250$, we have $\mathcal{T}_{t>250} \equiv \mathcal{M}_2$. That is, we change the model at iteration $t^* = 250$.

We apply MAPF with $N = 10^5$ total number of particles, $\epsilon = 0.1$, considering $\widehat{ESS}(\bar{\gamma}) = \frac{1}{\sum_{k=1}^{K} \sum_{m=1}^{M_{k,t}} \left(\bar{\gamma}_{k,t}^{(m)}\right)^2}$, and refreshing window $T_V = 125$ (clearly, $K = 2$). Given the previous assumptions, we have averaged the results over $10^4$ independent simulations, where the hidden states and the data are generated at each run, according to the model $\mathcal{T}_t$. Figure 5(a) shows the true and the estimated sequence of states in one specific run. Figure 5(b) depicts the evolution of the number of particles of each filters, $M_{k,t}$, $k = 1, 2$, as function of the the iterations $t$, in one particular run. Note that, the algorithm quickly detects the true model and assigns adequately all the computational effort to the filter addressing the true model, at the corresponding iteration $t$. Figure 5(b) also illustrates that MAPF recovers quickly after a refreshing step, each $T_V = 125$ iterations. We have also computed the Mean Square Error (MSE) in the estimation of $x_{1:T}$, and then averaged it over the $10^4$ independent runs. We compare the MSE obtained by MAPF with different unique particle filters with $N = 10^5$ particles: one PF considers the true model (best case), a second PF considers always $\mathcal{M}_1$, the third PF always deals with $\mathcal{M}_2$ and the last one addresses always the wrong model, i.e., $\mathcal{M}_2$ for $t \leq 250$ and $\mathcal{M}_1$ for $t > 250$. The results are shown in Table 3. MAPF provides an MSE virtually identical to the MSE of the best case, obtained using a unique PF addressing always the true model.

We also test MAPF considering different values of $T_V$ and using the adaptive procedure for selecting the refreshing steps described in Section 5, with $p_r = 0.1$. As suggested in Section 5, we use a mix strategy forcing also the refreshing procedure at the iteration 350, 410 and 450

| PF true model | MAPF ($T_V = 125$) | PF - $\mathcal{M}_1$ | PF - $\mathcal{M}_2$ | PF wrong model |
|---|---|---|---|---|
| 6.64 | 6.91 | 95.09 | 106.21 | 115.44 |

Table 3: MSE of MAPF in estimation of $x_{1:T}$, compared with the MSE of different unique PFs using the same number of particles $N = 10^5$.

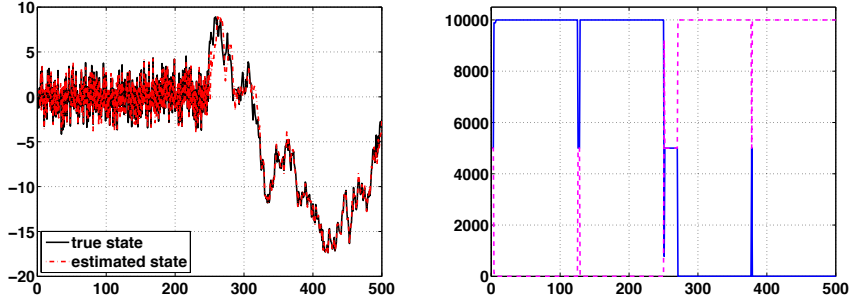

(a) True and estimated path in one run. (b) Num. of particles, $M_{1,t}$, $M_{2,t}$, as function of $t$.

Figure 5: (a) True (solid line) and estimated (dashed) sequence of states $x_{1:T}$ in one specific run, obtained by MAPF. (b) Evolution of the numbers of particles $M_{1,t}$, $M_{2,t}$ as function of the iterations $t$, in one specific run. A refreshing step is applied at each $T_V = 125$ iterations.

(note that this time steps are far from the time change of the model at the iteration 250). The results are shown in Table 4. We can observe that, in any case, MAPF provides MSE values much closer to the a single PF considering a unique true model (with an MSE of 6.64) than a single PF considering only $\mathcal{M}_1$ or $\mathcal{M}_2$ (with MSEs of 95.09 and 106.21, respectively). The good results obtained with $T_V \leq 50$ show that the application of the global resampling can be beneficial improving the global estimation (note that the MSE is small even if $T_V$ is not a divisor of 250). The results obtained with a $T_V > 250$ show the ability of MAPF of recovering the right model after a previous time change. Finally, observe that the automatic adaptive procedure is a reasonable option as robust tuning, i.e., when a specific selection by user (in advance) of $T_V$ is problematic.

| $T_V$ | 17 | 20 | 35 | 50 | 100 | 125 | 250 | 260 | 300 | adaptive |
|---|---|---|---|---|---|---|---|---|---|---|
| MSE | 6.95 | 6.80 | 7.12 | 6.84 | 15.78 | 6.91 | 6.88 | 7.08 | 21.68 | 8.03 |

Table 4: MSE of MAPF in estimation of $x_{1:T}$ with different values of $T_V$ and using the adaptive procedure in Section 5 with $p_r = 0.1$ ($N = 10^5$).

## 6.2 Parameter selection

Consider the system of equations defining different models, with $x_{k,t} \in \mathbb{R}$,

$$\mathcal{M}_k : \begin{cases} x_{k,t} = a_k |x_{k,t-1}| + v_{k,t} \\ y_t = b_k \log(x_{k,t}^2) + u_{k,t} \end{cases}, \qquad t = 1, \ldots, T. \tag{48}$$

where $v_{k,t} \sim \mathcal{N}(0, \sigma_{1,k}^2)$ and $u_{k,t} \sim \mathcal{N}(0, \sigma_{2,k}^2)$. We consider that true model $\mathcal{T}$ coincides with $\mathcal{M}_K$ with parameters $a_K = b_K = 1$, and $\sigma_{1,k} = \sigma_{2,k} = 1$. Thus, at each run, all the data $y_{1:T}$ are generated according to $\mathcal{T}$. We consider 3 different experimental settings:

- **General setup (S1):** in this case, each model $\mathcal{M}_k$ has both different dynamic equations and different likelihood functions. Specifically, we consider a grid of parameters,

$$a_k = \frac{k}{K}, \qquad k = 1, \ldots, K. \tag{49}$$

Note that $a_K = 1$. The first $K-1$ values of $\sigma_{1,k}$ are chosen randomly at each run, more precisely, $\sigma_{i,k} \sim \mathcal{U}([0.1, 10])$ for $i = 1, 2$ and $k = 1, \ldots, K-1$. Moreover, we have another grid for $b_k$,

$$b_k = \frac{1}{3} + \frac{10(k-1)}{K}, \qquad k = 1, \ldots, K-1, \tag{50}$$

and $b_K = 1$.
- **Common likelihood function (S2):** the parameters $a_k$ are selected are in Eq. (49), and again $\sigma_{1,k} \sim \mathcal{U}([0.1, 10])$, whereas in this case, we set $b_k = 1$, $\sigma_{2,k} = 1$ for all $k = 1, \ldots, K$. In other words, the models share the same likelihood function.
- **Common dynamic model (S3):** in this case, $a_k = 1$ and $\sigma_{1,k} = 1$ for all $k = 1, \ldots, K$, whereas the $b_k$s are set as in Eq. (50) (with $b_K = 1$) and $\sigma_{2,k} \sim \mathcal{U}([0.1, 10])$.

We apply MAPF performing 500 independent runs for every scenario, S1, S2 and S3. In each experiment, new hidden sequences and observed data are generated from the true model $\mathcal{T} \equiv \mathcal{M}_K$. In all cases, we set $T = 500$, the number of particles is $N = 10^5$, and the resampling parameter is $\epsilon = 0.1$. We also test different values of $T_V \in \{20, 40, 100, T+1\}$ (the case $T_V = T+1$ corresponds to a "non-refreshing" setup) and number of considered models $K \in \{5, 20, 50, 100\}$. Table 5 provides the percentage of the perfect match between the estimated model obtained by the maximum a posteriori (MAP) estimator and the true model. This percentage is averaged over the $T$ iterations, at each run. The results show that MAPF is able to detect the true model in different scenarios, even with very frequent refreshing steps. Figures 6 show the variable number of particles (in one specific run) of $K = 50$ filters as function of different iterations $t$ in different experimental setup. These figures confirm that MAPF is able to recover quickly the true model after several refreshing steps. Finally, considering $K = 5$, the numerical simulations also show that MAPF obtains a MSE approximately 10 times smaller than a unique filter with $N = 10^5$ particles employed for targeting a wrong model, without taking into account "catastrophic" runs when the filter is completely lost.

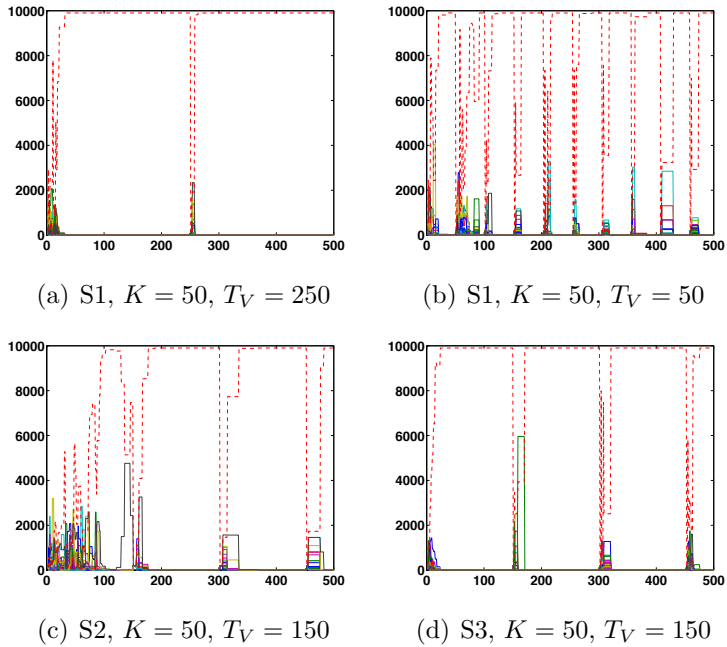

Figure 6: Number of particles $M_{k,t}$ (with $K = 50$ filters and $N = 10^5$) as function of the iterations $t$, in one specific run. The number of particles of the filter corresponding to the true model is depicted with a dashed line. Figures (a) and (b) correspond to the setup S1 and $T_V = 50, 150$, respectively; Figures (c) and (d) correspond to the setup S2 and S3 with $T_V = 150$, in both cases.

## 6.3 Real Data: Urban Mobility

In this section, we conduct a study on real-world data involving smartphone-based tracking. The knowledge of the current position is useful for many applications, for example to offer real-time location-based services such as advising the traveller when it is time to change bus, or providing other kinds of services such as warnings about traffic interruptions; or, simply collecting accurate travel paths for later analysis.

Location is obtained by the device's GPS receiver. However, this causes a relatively strong drain on the battery, so infrequent use of the GPS receiver is strongly desired. Furthermore, GPS reception does not work in some parts of the transport system, for example in the metro system (when it is underground) and on the train (due to unfavorable reception conditions caused by the metal cabin and overhead electrical lines). Hence, in this case, other kind of sensors should incorporated to the sensor network. A volunteer group of researchers collected GPS measurements, recorded continuously during two weeks by their Android smartphones running the CONTEXTLOGGER application[13]. To obtain a relatively accurate ground truth, we obtained readings every 10 seconds. In correspondence with our recorded data, we considered the time step as 10 seconds (so that the refreshing time $T_V$ must take into account this consideration). We assume a linear observation model $\mathbf{y}_t = \mathbf{x}_t + \mathbf{r}_t$ where $\mathbf{x}_t$ is the current position and $\mathbf{r}_t$ is a

---

[13]See http://contextlogger.org/.

Table 5: Percentage (averaged over the $T$ iterations and 500 independent runs) of the perfect match between the estimated model and the true model, in different settings (S1, S2 and S3) and number of models $K$.

(a) Without refreshing

| $K$ | 5 | 20 | 50 | 100 |
|---|---|---|---|---|
| S1 | 98.55 | 98.60 | 98.60 | 98.59 |
| S2 | 98.48 | 97.53 | 97.49 | 97.26 |
| S3 | 98.50 | 98.57 | 98.51 | 98.58 |

(b) With refreshing $T_V = 20$

| $K$ | 5 | 20 | 50 | 100 |
|---|---|---|---|---|
| S1 | 62.54 | 91.12 | 91.80 | 91.68 |
| S2 | 62.06 | 79.04 | 81.99 | 81.26 |
| S3 | 64.54 | 91.04 | 91.90 | 92.40 |

(c) With refreshing $T_V = 40$

| $K$ | 5 | 20 | 50 | 100 |
|---|---|---|---|---|
| S1 | 71.81 | 94.78 | 95.00 | 95.50 |
| S2 | 82.65 | 89.37 | 89.65 | 90.36 |
| S3 | 82.78 | 94.78 | 95.42 | 95.36 |

(d) With refreshing $T_V = 100$

| $K$ | 5 | 20 | 50 | 100 |
|---|---|---|---|---|
| S1 | 78.47 | 97.98 | 98.08 | 98.02 |
| S2 | 92.46 | 95.83 | 95.68 | 96.15 |
| S3 | 94.46 | 97.76 | 98.06 | 97.93 |

Gaussian noise with a standard deviation error of ±10.62 meters, as recorded in the true data (accuracy is recorded on each datum). Geographical coordinates of interactions (crossings) and bus stops can be obtained from OPENSTREETMAPS,[14] and, in the case of Helsinki which we have considered, from the urban planner[15]. We plot all stops in the Helsinki region in Fig. 7. There are several possibilities for modeling the different possible modes in urban mobility problem. Here, we provide some very simple and efficient possibilities, which can be easily implemented within a commercial application:

**Dynamic Model for Train-Bus.** Let us define the subset $\mathcal{S} \subseteq \mathbb{R}^2$ as a piecewise linear approximation of the route of the corresponding train or bus. Namely, the train tracks or the bus routes can be interpreted geometrically as curves embedded in $\mathbb{R}^2$. One set $\mathcal{S}$ is an approximation of one of these curves, obtained by a piecewise linear interpolation the corresponding route. An example is shown in Figure 8(a). We consider the following equation,

$$\mathbf{x}_t = \mathbf{x}_{t-1} + \mathbf{h}_t(\mathcal{S}), \tag{51}$$

where, $\mathbf{x}_t \in \mathbb{R}^2$, and $\mathbf{h}_t(\mathcal{S})$ is random Gaussian perturbation depending on the subset $\mathcal{S} \subseteq \mathbb{R}^2$. In this case, $\mathbf{x}_t \sim \mathcal{N}(\mathbf{x}_{t-1}, \mathbf{\Sigma}_t(\mathcal{S}))$, where the covariance matrix depends on $\mathcal{S}$: the slope of the eigenvector associated to the greatest eigenvalue $\lambda_1$ is the same of the current linear piece of $\mathcal{S}$ (see Figure 8(a)). Moreover, denoting as $\lambda_2$ the second eigenvalue, we design the covariance matrix $\mathbf{\Sigma}_t(\mathcal{S})$ in order to have $\lambda_1 >> \lambda_2$, i.e., the generated particles are highly correlated in the direction of the route. Clearly, the different trains, trams or buses are discriminated by the different routes, i.e, the set $\mathcal{S}$. Furthermore, the covariance matrix $\mathbf{\Sigma}_t(\mathcal{S})$ can also contain other kinds of information, such as the velocity; indeed, it is possible to choose properly the values of the two eigenvalues $\lambda_1, \lambda_2$. One possibility to improve the particle generation is to apply the rejection sampling principle [29, 33] for discarded the samples outside the route $\mathcal{S}$ (in this case, the noise is a truncated Gaussian pdf, restricted within $\mathcal{S}$). The values $\lambda_1, \lambda_2$ are tuned using a Least Squares pre-processing according to the data, depending go the specific vehicle (train, tram, bus) and route.

---

[14]See http://www.openstreetmap.org
[15]See http://hsl.fi

**Dynamic Model for Car, Cycling and Walk.** In this case, we consider a simple model of type

$$\mathbf{x}_t = \begin{cases} \mathbf{x}_{t-1} + b_1 \mathbf{v}_t, & \text{with prob. } \frac{1}{3}, \\ \mathbf{x}_{t-1} + b_2 \mathbf{v}_t, & \text{with prob. } \frac{1}{3}, \\ \mathbf{x}_{t-1} + b_3 \mathbf{v}_t, & \text{with prob. } \frac{1}{3}, \end{cases} \tag{52}$$

where $\mathbf{v}_t \sim \mathcal{N}(\mathbf{0}, \mathbf{I})$ and $b_1 > b_2 > b_3 = 0.05$ are scalar parameters representing fast, moderate, and slow movements, respectively. The parameter $b_3 = 0.05$ corresponds to waiting in a bus stop or waiting for a traffic light switch. In our experiments, in a city environment, we have set $b_1 = 3.1$, $b_2 = 1.6$, for motorized vehicles, whereas $b_1 = 2$, $b_2 = 0.8$ for cycling and $b_1 = 1$, $b_2 = 0.3$ for modeling walkers. However, we note that the results are not strongly conditionated on these choices.

Our experience suggest of using $50 \leq T_V \leq 200$ seconds. With this range of values $50 \leq T_V \leq 200$, we obtain a percentage greater of 82% in the estimation of the true modality, in the experiments considering the collected data. Figure 8(b) shows the results of a specific test taking into account 4 different modalities. Furthermore, we obtain an averaged MSE of 3.14 meters in the tracking estimation, considering all the different modalities. With the automatic adaptive procedure in Section 5 ($p_r = 0.1$), we obtain an MSE of 3.27 meters. Considering only buses and trains, we obtain an MSE of 1.17 meters (1.38 with the automatic-adaptive procedure). We have also implemented and tested the *multiple model bootstrap particle filter* proposed in [31] considering a uniform transition matrix for the underlying Markov process, required for modeling the transitions among the different modalities. Considering all the modalities, we have in this case an MSE of 4.76 meters, whereas considering only buses and trains an MSE of 1.59 meters. In any case, these MSEs are greater than the MSEs obtained by MAPF. The multiple model bootstrap particle filter probably requires a more accurate tuning of all the entries of the transition matrix of the the underlying Markov process (they are always more than 1, unlike in MAPF where we have only $T_V$).

# 7 Conclusions

We have designed an interacting parallel sequential Monte Carlo scheme for inference in state space models and online model selection. The parallel particle filters collaborate for providing a global efficient estimate of the hidden states and an approximation of the probability of the models given the received measurements. The exchange of information among the filters takes place through the adaptation of the numbers of particles of each filter. A exhaustive theoretical derivation has been provided. The proposed technique has been applied successfully in different experimental scenarios, including a real data experiment for the mode detection in a urban mobility problem.

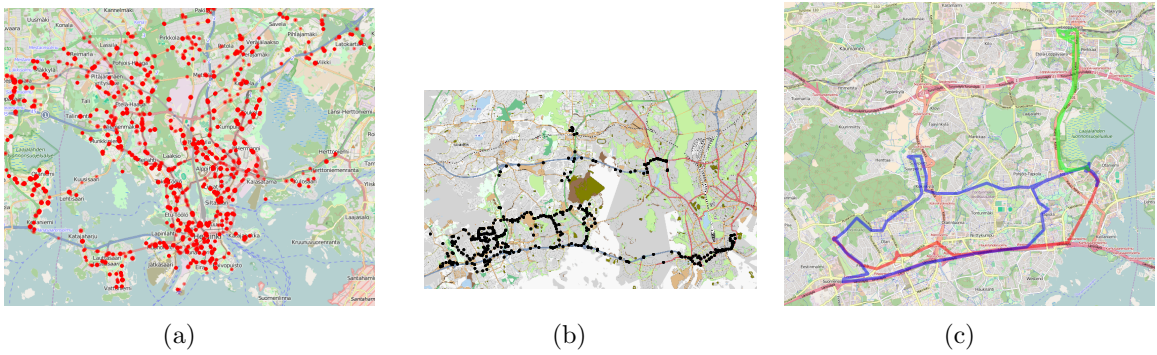

Figure 7: **(a)** All bus stops in Helsinki region (left). **(b)** All intersections crossed during a 10-day period by one participant. **(c)** A small section of trajectories labeled under different transportation modes. Blue is car, red is bus, and green is cycling.

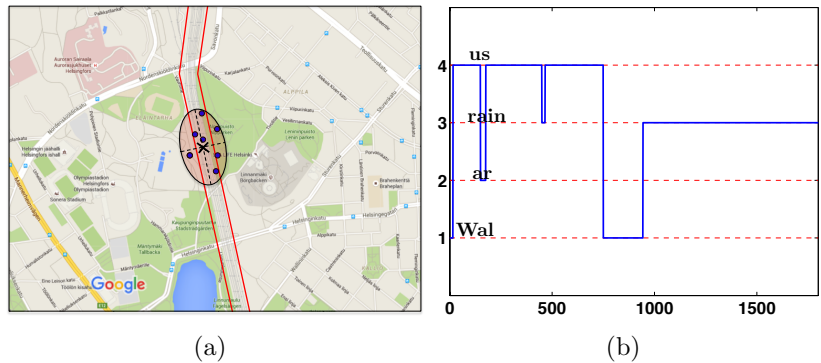

Figure 8: **(a)** Example of generation of particles (shown as circles) using the dynamic model of type in Eq. (51). **(b)** Detection of transition between Bus 55, Walk and Train (MAP estimation) at the Helsinki central railway station, considering 4 possible modalities with real data ($T \approx 1900$ sec and $T_V = 150$ sec).

# Acknowledgement

This work has been supported by the Grant 2014/23160-6 of São Paulo Research Foundation (FAPESP), the Grant 305361/2013-3 of National Council for Scientific and Technological Development (CNPq), and the Aalto University AEF research programme http://energyefficiency.aalto.fi/en/.

# A  Recursive formulas for sequential inference

For simplifying the notation, in this section, we consider working only with a unique model $\mathcal{M}_1$, so that $p(\mathbf{x}_{1:t}, \mathbf{y}_{1:t}) = p(\mathbf{x}_{1:t}, \mathbf{y}_{1:t}|\mathcal{M}_1)$. Thus, we can write easily the following recursive formula for the joint pdf $p(\mathbf{x}_{1:t}, \mathbf{y}_{1:t})$ [13, 15],

$$p(\mathbf{x}_{1:t}, \mathbf{y}_{1:t}) = \ell_t(\mathbf{y}_t|\mathbf{x}_t) q_t(\mathbf{x}_t|\mathbf{x}_{t-1}) \, p(\mathbf{x}_{1:t-1}, \mathbf{y}_{1:t-1}).$$

However, we are interested in a similar recursive expression which involves the posterior $p(\mathbf{x}_{1:t}|\mathbf{y}_{1:t}) = \frac{p(\mathbf{x}_{1:t}, \mathbf{y}_{1:t})}{p(\mathbf{y}_{1:t})}$. Thus, starting be the definition and replacing the expression above, we obtain

$$p(\mathbf{x}_{1:t}|\mathbf{y}_{1:t}) = \frac{\ell_t(\mathbf{y}_t|\mathbf{x}_t) q_t(\mathbf{x}_t|\mathbf{x}_{t-1})}{p(\mathbf{y}_{1:t})} p(\mathbf{x}_{1:t-1}, \mathbf{y}_{1:t-1}). \tag{53}$$

Replacing $p(\mathbf{x}_{1:t-1}, \mathbf{y}_{1:t-1}) = p(\mathbf{x}_{1:t-1}|\mathbf{y}_{1:t-1}) p(\mathbf{y}_{1:t-1})$, we have

$$p(\mathbf{x}_{1:t}|\mathbf{y}_{1:t}) = \ell_t(\mathbf{y}_t|\mathbf{x}_t) q_t(\mathbf{x}_t|\mathbf{x}_{t-1}) \frac{p(\mathbf{y}_{1:t-1})}{p(\mathbf{y}_{1:t})} p(\mathbf{x}_{1:t-1}|\mathbf{y}_{1:t-1}),$$

and since we can write $p(\mathbf{y}_{1:t}) = p(\mathbf{y}_t|\mathbf{y}_{1:t-1})p(\mathbf{y}_{1:t-1})$, finally we obtain

$$
\begin{aligned}
p(\mathbf{x}_{1:t}|\mathbf{y}_{1:t}) &= [\ell_t(\mathbf{y}_t|\mathbf{x}_t)q_t(\mathbf{x}_t|\mathbf{x}_{t-1})] \frac{p(\mathbf{y}_{1:t-1})}{p(\mathbf{y}_t|\mathbf{y}_{1:t-1})p(\mathbf{y}_{1:t-1})} p(\mathbf{x}_{1:t-1}|\mathbf{y}_{1:t-1}), \\
&= \frac{\ell_t(\mathbf{y}_t|\mathbf{x}_t)q_t(\mathbf{x}_t|\mathbf{x}_{t-1})}{p(\mathbf{y}_t|\mathbf{y}_{1:t-1})} p(\mathbf{x}_{1:t-1}|\mathbf{y}_{1:t-1}).
\end{aligned}
$$

The last expression involves the posterior at the $t$-th iteration, $p(\mathbf{x}_{1:t}|\mathbf{y}_{1:t})$, as function of the posterior at the $t-1$-th iteration, $p(\mathbf{x}_{1:t-1}|\mathbf{y}_{1:t-1})$.

# B  Derivation of the estimator $\widetilde{Z}_t$

For simplicity, we again assume working only with a unique model $\mathcal{M}_1$, so that $K = 1$ and $p(\mathbf{x}_{1:t}, \mathbf{y}_{1:t}) = p(\mathbf{x}_{1:t}, \mathbf{y}_{1:t}, \mathcal{M}_1) = p(\mathbf{x}_{1:t}, \mathbf{y}_{1:t}|\mathcal{M}_1)$. Within the SIS framework, there are two possible formulations of the estimator of $Z$, i.e., $\widehat{Z}$ in Eq. (27) and $\widetilde{Z}$ given in Eq. (28). The alternative formulation $\widetilde{Z}$ in Eq. (28) can be derived as follows [15]. Recall that

$$
Z_t = \int_{\mathcal{X}_{1:t}} p(\mathbf{x}_{1:t}, \mathbf{y}_{1:t}) d\mathbf{x}_{1:t} \approx \widehat{Z}_t = \frac{1}{M} \sum_{m=1}^{M} w_t^{(m)},
$$

and consider the following integral

$$
\begin{aligned}
\int_{\mathcal{X}_{1:t}} \ell_t(\mathbf{y}_t|\mathbf{x}_t)q_t(\mathbf{x}_t|\mathbf{x}_{t-1})p(\mathbf{x}_{1:t-1}|\mathbf{y}_{1:t-1})d\mathbf{x}_{1:t} &= \frac{1}{p(\mathbf{y}_{1:t-1})} \int_{\mathcal{X}_{1:t}} \ell_t(\mathbf{y}_t|\mathbf{x}_t)q_t(\mathbf{x}_t|\mathbf{x}_{t-1})p(\mathbf{x}_{1:t-1}, \mathbf{y}_{1:t-1})d\mathbf{x}_{1:t}, \\
&= \frac{1}{p(\mathbf{y}_{1:t-1})} \int_{\mathcal{X}_{1:t}} p(\mathbf{x}_{1:t}, \mathbf{y}_{1:t}) d\mathbf{x}_{1:t} \\
&= \frac{p(\mathbf{y}_{1:t})}{p(\mathbf{y}_{1:t-1})} = \frac{Z_t}{Z_{t-1}} = p(\mathbf{y}_t|\mathbf{y}_{1:t-1}).
\end{aligned}
$$

where in the last expression we have used $Z_t = p(\mathbf{y}_{1:t})$ and $p(\mathbf{y}_{1:t}) = p(\mathbf{y}_t|\mathbf{y}_{1:t-1})p(\mathbf{y}_{1:t-1})$. Summarizing, we have obtained

$$
\int_{\mathcal{X}_{1:t}} \ell_t(\mathbf{y}_t|\mathbf{x}_t)q_t(\mathbf{x}_t|\mathbf{x}_{t-1})p(\mathbf{x}_{1:t-1}|\mathbf{y}_{1:t-1})d\mathbf{x}_{1:t} = \frac{Z_t}{Z_{t-1}}. \tag{54}
$$

Now we replace above $p(\mathbf{x}_{1:t-1}|\mathbf{y}_{1:t-1})$ the particle approximation $\widehat{p}(\mathbf{x}_{1:t-1}|\mathbf{y}_{1:t-1}) = \sum_{m=1}^{M} \bar{w}_{t-1}^{(m)} \delta(\mathbf{x}_{1:t-1} - \mathbf{x}_{1:t-1}^{(m)})$, so we can write

$$
\int_{\mathcal{X}_{1:t}} \ell_t(\mathbf{y}_t|\mathbf{x}_t)q_t(\mathbf{x}_t|\mathbf{x}_{t-1})\widehat{p}(\mathbf{x}_{1:t-1}|\mathbf{y}_{1:t-1})d\mathbf{x}_{1:t} \sum_{m=1}^{M} \bar{w}_{t-1}^{(m)} \int_{\mathcal{X}_t} \ell_t(\mathbf{y}_t|\mathbf{x}_t)q_t(\mathbf{x}_t|\mathbf{x}_{t-1}^{(m)})d\mathbf{x}_t, \approx \frac{Z_t}{Z_{t-1}}. \tag{55}
$$

Moreover, approximating the $M$ integrals $\int_{\mathcal{X}_t} \ell_t(\mathbf{y}_t|\mathbf{x}_t)q_t(\mathbf{x}_t|\mathbf{x}_{t-1}^{(m)})d\mathbf{x}_t$ via Monte Carlo using *only one sample*, $\mathbf{x}_t^{(m)} \sim \phi_t(\mathbf{x}_t^{(m)}|\mathbf{x}_{1:t-1}^{(m)})$, for each one,

$$\sum_{m=1}^{M} \bar{w}_{t-1}^{(m)} \int_{\mathcal{X}_t} \ell_t(\mathbf{y}_t|\mathbf{x}_t)q_t(\mathbf{x}_t|\mathbf{x}_{t-1}^{(m)})d\mathbf{x}_t \approx \sum_{m=1}^{M} \bar{w}_{t-1}^{(m)} \frac{\ell_t(\mathbf{y}_t|\mathbf{x}_t^{(m)})q_t(\mathbf{x}_t^{(m)}|\mathbf{x}_{t-1}^{(m)})}{\phi_t(\mathbf{x}_t^{(m)}|\mathbf{x}_{1:t-1}^{(m)})},$$
$$= \sum_{m=1}^{M} \bar{w}_{t-1}^{(m)} \lambda_t^{(m)} \approx \frac{Z_t}{Z_{t-1}}. \qquad (56)$$

**Alternative derivation.** Note that we can also deduce the past expression as following

$$\begin{aligned}
\sum_{m=1}^{M} \bar{w}_{t-1}^{(m)} \lambda_t^{(m)} &= \frac{1}{\sum_{m=1}^{M} w_{t-1}^{(m)}} \sum_{m=1}^{M} w_{t-1}^{(m)} \lambda_t^{(m)} \\
&= \frac{1}{\sum_{m=1}^{M} w_{t-1}^{(m)}} \sum_{m=1}^{M} w_t^{(m)} = \frac{\frac{1}{M}\sum_{m=1}^{M} w_t^{(m)}}{\frac{1}{M}\sum_{m=1}^{M} w_{t-1}^{(m)}} \\
&= \frac{\widehat{Z}_t}{\widehat{Z}_{t-1}} \approx \frac{Z_t}{Z_{t-1}}.
\end{aligned}$$

**Equivalence with $\widehat{Z}_t$.** Setting $\widehat{Z}_0 = 1$, we can obtain that

$$\widetilde{Z}_t = \prod_{\tau=1}^{t} \left[ \sum_{m=1}^{M} \bar{w}_{\tau-1}^{(m)} \lambda_\tau^{(m)} \right] = \widehat{Z}_1 \frac{\widehat{Z}_2}{\widehat{Z}_1} \cdots \frac{\widehat{Z}_{t-1}}{\widehat{Z}_{t-2}} \frac{\widehat{Z}_t}{\widehat{Z}_{t-1}} = \widehat{Z}_t \approx Z, \qquad (57)$$

namely, the estimator in Eq. (28) is exactly equivalent the estimator (27).

## B.1 Application of resampling.

Consider again to approximate the integral in Eq. (55) via importance sampling. In this case, we assume to draw independent samples $\mathbf{x}_t^{(1)}, \ldots, \mathbf{x}_t^{(M)}$ from the a different proposal pdf $\varphi(\mathbf{x}_{1:t})$, defined as

$$\varphi(\mathbf{x}_{1:t}) = \phi_t(\mathbf{x}_t|\mathbf{x}_{1:t-1})\widehat{p}(\mathbf{x}_{1:t-1}|\mathbf{y}_{1:t-1}) = \sum_{m=1}^{M} \bar{w}_{t-1}^{(m)} \phi_t(\mathbf{x}_t|\mathbf{x}_{1:t-1}^{(m)}).$$

Note that, this is equivalent to apply a resampling at the $(t-1)$-th iteration. Thus, we can write

$$\int_{\mathcal{X}_{1:t}} \ell_t(\mathbf{y}_t|\mathbf{x}_t)q_t(\mathbf{x}_t|\mathbf{x}_{t-1})\widehat{p}(\mathbf{x}_{1:t-1}|\mathbf{y}_{1:t-1})d\mathbf{x}_{1:t} \approx$$
$$\frac{1}{M}\sum_{m=1}^{M} \frac{\ell_t(\mathbf{y}_t|\mathbf{x}_t^{(m)})q_t(\mathbf{x}_t^{(m)}|\mathbf{x}_{t-1}^{(m)})}{\phi_t(\mathbf{x}_t^{(m)}|\mathbf{x}_{1:t-1}^{(m)})} = \frac{1}{M}\sum_{m=1}^{M} \lambda_t^{(m)} \approx \frac{Z_t}{Z_{t-1}}. \qquad (58)$$

where $\mathbf{x}_t^{(m)} \sim \phi_t(\mathbf{x}_t|\mathbf{x}_{1:t-1})\widehat{p}(\mathbf{x}_{1:t-1}|\mathbf{y}_{1:t-1})$, with $m = 1, \ldots, M$. Recalling the definition of $\xi_t^{(m)}$,

$$\xi_t^{(m)} = \begin{cases} w_t^{(m)}, & \text{without resampling at the } t\text{-th iteration,} \\ \widehat{Z}_t, & \text{after resampling at the } t\text{-th iteration,} \end{cases} \quad (59)$$

we can ensure that $\widehat{Z}_t = \sum_{m=1}^M \xi_t^{(m)} \lambda_t^{(m)}$, is still a valid estimator of $Z$. When no resampling is performed, $\xi_{t-1}^{(m)} = w_{t-1}^{(m)}$, we come back to the standard IS estimator of $Z$. When the resampling is applied, we have $\xi_{t-1}^{(m)} = \widehat{Z}_{t-1}$, then $\widehat{Z}_t = \frac{\widehat{Z}_{t-1}}{M} \sum_{m=1}^M \lambda_t^{(m)}$. Since $\widehat{\frac{Z_t}{Z_{t-1}}} = \frac{1}{M} \sum_{m=1}^M \lambda_t^{(m)}$, we obtain

$$\widehat{Z}_t = \widehat{Z}_{t-1} \left[ \frac{1}{M} \sum_{m=1}^M \lambda_t^{(m)} \right] = \widehat{Z}_{t-1} \widehat{\frac{Z_t}{Z_{t-1}}} \approx Z.$$

Finally, note that $\widehat{Z}_t$ and $\widetilde{Z}_t = \prod_{t=1}^T \left[ \sum_{m=1}^M \bar{\xi}_{t-1}^{(m)} \lambda_t^{(m)} \right]$ are two equivalent formulations of the same estimator. It can be shown exactly as described above for the SIS framework, replacing $w_t^{(m)}$ with $\xi_t^{(m)}$. Furthermore, if the resampling is applied at each iteration, observe that both estimators become

$$\widetilde{Z}_t = \prod_{\tau=1}^t \left[ \frac{1}{M} \sum_{j=1}^M \lambda_\tau^{(j)} \right], \quad (60)$$

and

$$\widehat{Z}_t = \widehat{Z}_{t-1} \left[ \frac{1}{M} \sum_{m=1}^M \lambda_t^{(m)} \right] = \prod_{\tau=1}^t \left[ \frac{1}{M} \sum_{j=1}^M \lambda_\tau^{(j)} \right]. \quad (61)$$

Note that, with respect to the estimator in Eq. (27) (for SIS, i.e., without resampling), the operations of product and sum are inverted.

## C  Further considerations about MAPF

The validity of the MAPF scheme relies on each filter performs separately a proper SIR estimation of the hidden states using the normalized weights $\bar{w}_{k,t}^{(i_k)}$. Each filter also provides a consistent estimator $\widehat{Z}_{k,t}$ of the marginal likelihood. Then, this information is properly merged following the approximation of $\widehat{p}(\mathbf{x}|\mathbf{y})$ given in Eq. (15) or (20). The variable numbers of particles does not provide any theoretical issues, since we obtain always valid IS estimators (clearly, it affects the efficiency of these estimators) [16]. An important related observation is remarked below.

**Remark 5.** *Consider that, at the $t$-th iteration, the condition in Eq. (37) is satisfied. Namely,* **(a)** *the numbers of particles are updated and* **(b)** *the resampling applied. Considering jointly* **(a)** *and* **(b)**, *we can be interpreted that in MAPF we are drawing samples from the mixture in Eq. (20),*

$$\widehat{p}(\mathbf{x}_{1:t}|\mathbf{y}_{1:t}) = \sum_{k=1}^K \bar{\rho}_{k,t} \, \widehat{p}(\mathbf{x}_{1:t}|\mathbf{y}_{1:t}, \mathcal{M}_k),$$

*using the so-called deterministic mixture (DM) procedure [16]. The DM is performed through the adaptation of the number of particles.*

That is, the adaptation of number of particles can be seen as a way of selecting (i.e., using) more times one model than other. The previous remark suggests an alternative equivalent resampling scheme:

1. Set $i_k = 0$, for all $k = 1, \ldots, K$.

2. For $n = 1, \ldots, N$:

    (a) Select a model $k$ with probability $\bar{\rho}_{k,t}$, $k = 1, \ldots, K$.

    (b) Set $i_k = i_k + 1$ resample $\bar{\mathbf{x}}_{k,t}^{(i_k)} = \mathbf{x}_{k,t}^{(j)} \in \{\mathbf{x}_{k,t}^{(1)}, \ldots, \mathbf{x}_{k,t}^{(M_{k,t})}\}$ with probability $\bar{w}_{k,t}^{(j)}$, $j = 1, \ldots, M_{k,t}$.

3. Set $M_{k,t+1} = i_k$, for all $k = 1, \ldots, K$, so that $\sum_{k=1}^{K} M_{k,t+1} = N$.

Clearly, with the procedure above, some model could be completely discarded, with no particles assigned.